\documentclass[preprint,flushrt]{aastex}
\usepackage{natbib}
\usepackage{lineno}
\shorttitle{November 2010 {\it 3C454.3} giant flare}
\shortauthors{Escande et al.}

\title{{\sl Fermi} Gamma-ray Space Telescope Observations of the Gamma-ray Outburst from 3C454.3 in November 2010}

\author{
A.~A.~Abdo\altaffilmark{1}, 
M.~Ackermann\altaffilmark{2}, 
M.~Ajello\altaffilmark{2}, 
A.~Allafort\altaffilmark{2}, 
L.~Baldini\altaffilmark{3}, 
J.~Ballet\altaffilmark{4}, 
G.~Barbiellini\altaffilmark{5,6}, 
D.~Bastieri\altaffilmark{7,8}, 
R.~Bellazzini\altaffilmark{3}, 
B.~Berenji\altaffilmark{2}
% ...
}
\altaffiltext{1}{National Research Council Research Associate, National Academy of Sciences, Washington, DC 20001, resident at Naval Research Laboratory, Washington, DC 20375}
\altaffiltext{2}{W. W. Hansen Experimental Physics Laboratory, Kavli Institute for Particle Astrophysics and Cosmology, Department of Physics and SLAC National Accelerator Laboratory, Stanford University, Stanford, CA 94305, USA}
\altaffiltext{3}{Istituto Nazionale di Fisica Nucleare, Sezione di Pisa, I-56127 Pisa, Italy}
\altaffiltext{4}{Laboratoire AIM, CEA-IRFU/CNRS/Universit\'e Paris Diderot, Service d'Astrophysique, CEA Saclay, 91191 Gif sur Yvette, France}
\altaffiltext{5}{Istituto Nazionale di Fisica Nucleare, Sezione di Trieste, I-34127 Trieste, Italy}
\altaffiltext{6}{Dipartimento di Fisica, Universit\`a di Trieste, I-34127 Trieste, Italy}
\altaffiltext{7}{Istituto Nazionale di Fisica Nucleare, Sezione di Padova, I-35131 Padova, Italy}
\altaffiltext{8}{Dipartimento di Fisica ``G. Galilei", Universit\`a di Padova, I-35131 Padova, Italy}
% ...

\begin{abstract}
The flat-spectrum radio quasar  3C454.3 underwent an extraordinary  5-day $\gamma$-ray outburst in November 2010 when the daily flux measured with the {\sl Fermi} Large Area Telescope (LAT) at photon energies $E>100\:$MeV 
reached  ($66\pm2)\times10^{-6}$ph cm$^{-2}$s$^{-1}$. 
This is  a factor of 3 higher than its previous maximum flux recorded in December 2009 and $\gtrsim5$ times brighter than the Vela pulsar, which is normally the brightest source in the $\gamma$-ray sky. The 3-hr peak flux was (85$\pm$5)$\times$10$^{-6}$ph cm$^{-2}$s$^{-1}$, corresponding to an apparent isotropic luminosity  of
(2.1$\pm$0.2)$\times$10$^{50}$ erg s$^{-1}$, the highest ever recorded for a
blazar. In this paper, we investigate the features of this exceptional event in the $\gamma$-ray band of the 
{\sl Fermi}-LAT. In contrast to previous flares of the same source observed with the {\sl Fermi}-LAT, 
clear spectral changes are observed during the flare. 

\end{abstract}
\keywords{Galaxies: active}
\begin{document}

\section{Introduction}

The radio source 3C454.3, a well-known flat-spectrum radio quasar (FSRQ) at redshift $z=0.859$,  has shown remarkably high activity since 2000. 
 It has been particularly bright in the $\gamma$-ray band covered by AGILE and the {\sl Fermi}-LAT, reaching a daily record flux level F$[E>100\:MeV]$ ($F_{100}$ in units of $10^{-6}$ph cm$^{-2}$s$^{-1}$) of $22\pm1$
  in December 2009 \citep{Str10,2010ApJ...721.1383A}. This high flux allowed detailed analysis to be performed, making it the best-studied blazar in the GeV band. Gamma-ray variability on timescales as short as a few hours 
\citep{Tav10} and a flux-doubling time scale of less than 3$\:$hours for a short subflare on 2009 Dec 5 \citep[MJD55170, ][]{2010ApJ...721.1383A} have been reported. 
In the LAT energy band, 3C454.3 exhibits  a spectrum with a clear departure from a power-law distribution, 
characterized by a break around 2$\:$GeV \citep{Abdo_3C}. This is found to be a common feature of bright $\gamma$-ray FSRQs \citep{spec_an}. 
Little variation of the break energy and spectra for large differences in flux states was observed for 3C454.3 \citep{2010ApJ...721.1383A}.  
From $\gamma\gamma$-opacity constraints, a minimum Doppler factor $\delta_{min}\approx 13$ was derived 
from the flux variability time \citep{2010ApJ...721.1383A}, and highest-energy photon measurements, in accord with independent measurements of $\delta$ from superluminal motion observations \citep{Jor05}. 

In November 2010, the source displayed sustained activity at a flux of $F_{100}\approx10$ 
for several days before showing a fast rise to record levels of $F_{100}=55$, then rising to $F_{100}\approx80$ (as measured over 6 hr-long periods).
In this paper, the intraday variability and the associated spectral changes in the $\gamma$-ray band of 3C454.3 are studied and comparisons are made with the findings obtained from earlier major flares. In Section 2, observations and analysis of {\sl Fermi}-LAT data from 2010 September 1 to December 13 are presented. 
Results are presented in Section 3 and  discussion is given in Section 4. A flat $\Lambda$CDM cosmology with $H_0=71\:$km s$^{-1}$ Mpc$^{-1}$, $\Omega_m=0.27$ and $\Omega_\Lambda$=0.73  is used in this paper.

\section{Observations and analysis}

The analysis
performed for this paper is very similar to that reported in \cite{2010ApJ...721.1383A}, to which we refer for details. 
The data presented in this paper are restricted to the 100$\:$MeV--200$\:$GeV range and were collected from
MJD55440 (2010 September 1) to MJD55543 (2010 December 13) in survey mode.  

Spectral analyses were performed by fitting the spectra with  multiple different models over the whole
energy range covered by the LAT at  $E>100\:$MeV. The spectral forms considered are a broken power law (BPL, $N(E)=N_0
(E/E_\textit{break})^{-\Gamma_{i}}$, with $i=1$ if $E<E_\textit{break}$ and $i=2$ if  $E>E_\textit{break}$),  a log-parabola function (\mbox{$N(E)=N_0\:(E/E_{p})^{-\alpha-\beta\:\log(E/E_p)}$}, where $E_p$ is
fixed at 1$\:$GeV),  a power law with exponential cutoff function  (PLEC, \mbox{$N(E)=N_0\:(E/E_{p})^{-\Gamma}\exp(-E/E_{\textit{cutoff}})$}), and a PL model over equally spaced logarithmic
energy bins with $\Gamma$ kept constant and equal to the value fitted over the whole
range.

Source variability was investigated by producing light curves with various time binnings (3$\:$hours, 6$\:$hours, 1$\:$day, 1$\:$week) and over different energy ranges (E$>$100$\:$MeV, E$>$1$\:$GeV, E=0.1-1$\:$GeV). 
Although the actual spectral shape exhibits definite curvature, light curves were produced by modeling the spectra in each time bin as a simple power law (PL) over the considered energy range, since the statistical uncertainties on the power-law indices are smaller than those obtained from BPL fits. 
In order to minimize spurious correlations between integrated flux and  $\Gamma$, the fluxes $F_{E>E_0}$  were also computed above the ``decorrelation energy'' $E_0$ where this correlation is minimal. 
For the 2009 December and 2010 April flares,  $E_0$ was found to be 163$\:$MeV \citep{2010ApJ...721.1383A}. 
The same value has been used here for consistency.
The estimated systematic uncertainty on the flux is 10\% at 100$\:$MeV, 5\% at 500$\:$MeV and 20\% at 10$\:$GeV. The energy resolution is better than 10\% over the range of measured E$_\textit{break}$.

\section{Results}

{\bf Figure \ref{fig:light_curve_1} (top panel) showing the historical $F_{100}$ light curve  illustrates the spectacular rise in activity of 3C454.3 over the years. It is evident that the December 2010 outburst dwarfs any previously recorded flares. The second panel displays the light curves with time binnings of 1-day (open circles) and 1-week (filled circles) during the outburst period.} A 13-day-long plateau   precedes the 5-day-long flare, confirming the trend previously observed in the July 2008 and December 2009 flares, 
but it is longer in duration and higher in flux than those in previous flares \citep{2010ApJ...721.1383A}.  
The onset of this plateau is clearly accompanied by a weak but significant hardening of the spectrum ($\Gamma$ changes from $2.50\pm0.02$ to $2.32\pm0.03$), as observed on a weekly time scale in the bottom panel of Figure \ref{fig:light_curve_1}. 
The daily flux decreases by a factor of about 3 in 4 days at the end of the flare. The flare is followed by a slowly decaying activity around  $F_{100}=20$. Different time periods labeled pre-flare, plateau, flare and post-flare in  Figure \ref{fig:light_curve_1} are considered in the following.

\begin{figure}
  \centering
  \epsscale{1.0}
  \resizebox{14cm}{!}{\rotatebox[]{0}{\includegraphics{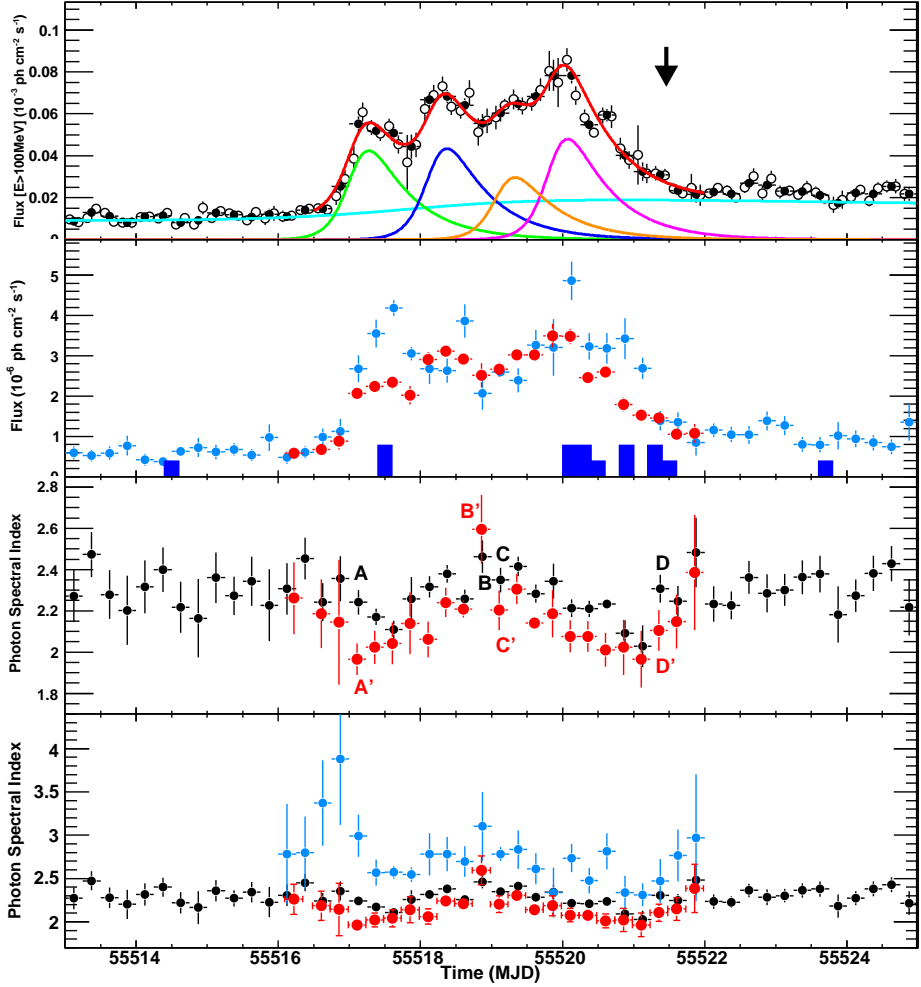}}}\\
  \caption{ Top panel: light curve of the flux above 100$\:$MeV. Open and filled symbols correspond to 3-hr and 6-hr averaged
  quantities respectively.  The lines correspond to the results of a five-component fit (four subflares and a slowly-varying background) as described in the text, using the 6-hr data. The arrow depicts the detection time of the 31-GeV photon.
Second panel: flux $F_{0.1-1{\rm GeV}}$, multiplied by a factor of 0.05 (red); flux $F_{1{\rm GeV}}$ (blue). The blue histogram represents the times of detection of $E>10$ GeV photons. 
Third panel: $\Gamma$ measured at E$>$100 MeV (black) and E=100$\:$MeV-1$\:$GeV (red). Bottom panel: $\Gamma$ measured at E$>$100$\:$MeV (black), E=100$\:$MeV-1$\:$GeV (red) and E$>$1$\:$GeV (blue). }
  \label{fig:light_curve_2}
  \end{figure}

The light curves  for $F_{100}$ with  6-hr and 3-hr time binnings (focusing on the flare period)  are shown in the upper panel of Figure \ref{fig:light_curve_2}. 
The F$[E>1\:$GeV] light curves with 6-hr and 3-hr time binnings are given in the second panel of Figure \ref{fig:light_curve_2}.
The corresponding evolution of $\Gamma$ is plotted in the 
third and bottom  panels. As can be seen, 
the major flare lasts for about 5$\:$days. In the $F_{100}$ light curves, 
it seems to comprise three to four subflares, with the flux peaking during the last one. 
The rise time of the MJD55516.5 flare is 12$\:$hours for a factor of 4 increase in flux 
(i.e., a doubling time of 6$\:$hours). This is shorter than  the previous fastest relative flux variation claimed in the GeV band for a major flare, which was from PKS 1502+106, 
when an increase by a factor of 3 in 12$\:$hours (i.e., a doubling time of 7.5$\:$hours)
was found \citep{2010ApJ...710..810A}. In the upper panel of Figure \ref{fig:light_curve_2}, a fit consisting of a slowly varying background and four faster temporally evolving components
 was performed between MJD55516.5 and MJD55522 for both the 6-hr and 3-hr light curves. 
Each component is assumed to be fit by a function of the form 
\begin{equation} 
F=2F_0(e^{(t_0-t)/T_r}+e^{(t-t_0)/T_f})^{-1} \;
\label{fit_function}
\end{equation}
\citep{Abdo_var}, where $T_r$ and $T_f$ are the rising and falling times, respectively, and $F_0$  is the flux at $t_0$ representing approximately the flare amplitude.  
With the rise time  $T_r$  set equal for all subflares, and likewise for the fall  time $T_f$,
we find that $T_r=4.5\pm1$hr and $T_f=15\pm2$hr gives a good fit to  both the 3-hr and 6-hr light curves. The highest-energy photon collected during the MJD55516-55522 period within the energy and inclination angle-dependent 95\% containment angle around the source position has $E_{max}=31\pm3\:$GeV  and was detected at MJD55521.46. 
Its detection time is depicted with an arrow in the upper panel of Figure \ref{fig:light_curve_2}.

Significant differences between light curves for fluxes in the E=0.1-1$\:$GeV ($F_{0.1-1\:{\rm GeV}}$) and 
E$>1\:$GeV  ($F_{1{\rm GeV}}$) ranges are observed (see the second panel of Figure \ref{fig:light_curve_2}). 
The peak of the first subflare occurs approximately 15$\:$hr after $F_{0.1-1{\rm GeV}}$ has leveled off, demonstrating clear spectral variability. 
This behavior is confirmed by a progressive decrease of $\Gamma$ (spectral hardening) from $\Gamma\:\approx\:2.35$ 
to  $\Gamma\:\approx\:2.1$ as the subflare develops beyond the 100$\:$MeV peak  (black points in the third panel of Figure \ref{fig:light_curve_2}). 
Overall, the  $F_{1{\rm GeV}}$ light curve shows sharper structures than the $F_{0.1-1{\rm GeV}}$ light curve. A 
clear difference is also observed in the decaying stage, with the high-energy component starting to fade away {\sl later} than the 
lower-energy component. In the lower two panels of Figure \ref{fig:light_curve_2}, the spectral indices obtained over the two restricted energy ranges are shown as well. 
As $F_{0.1-1\:{\rm GeV}}$ levels off in the first subflare while $F_{1\:{\rm GeV}}$ keeps rising, the $0.1-1\:{\rm GeV}$ index is fairly constant,
indicating that the hardening is limited to the range above 1$\:$GeV. This hardening is confirmed by the evolution of $\Gamma$ measured in the 
$>1\:$GeV range (blue points in the bottom panel in Figure \ref{fig:light_curve_2}).  

\begin{figure}[t]
\centering
\epsscale{1.0}
\plotone{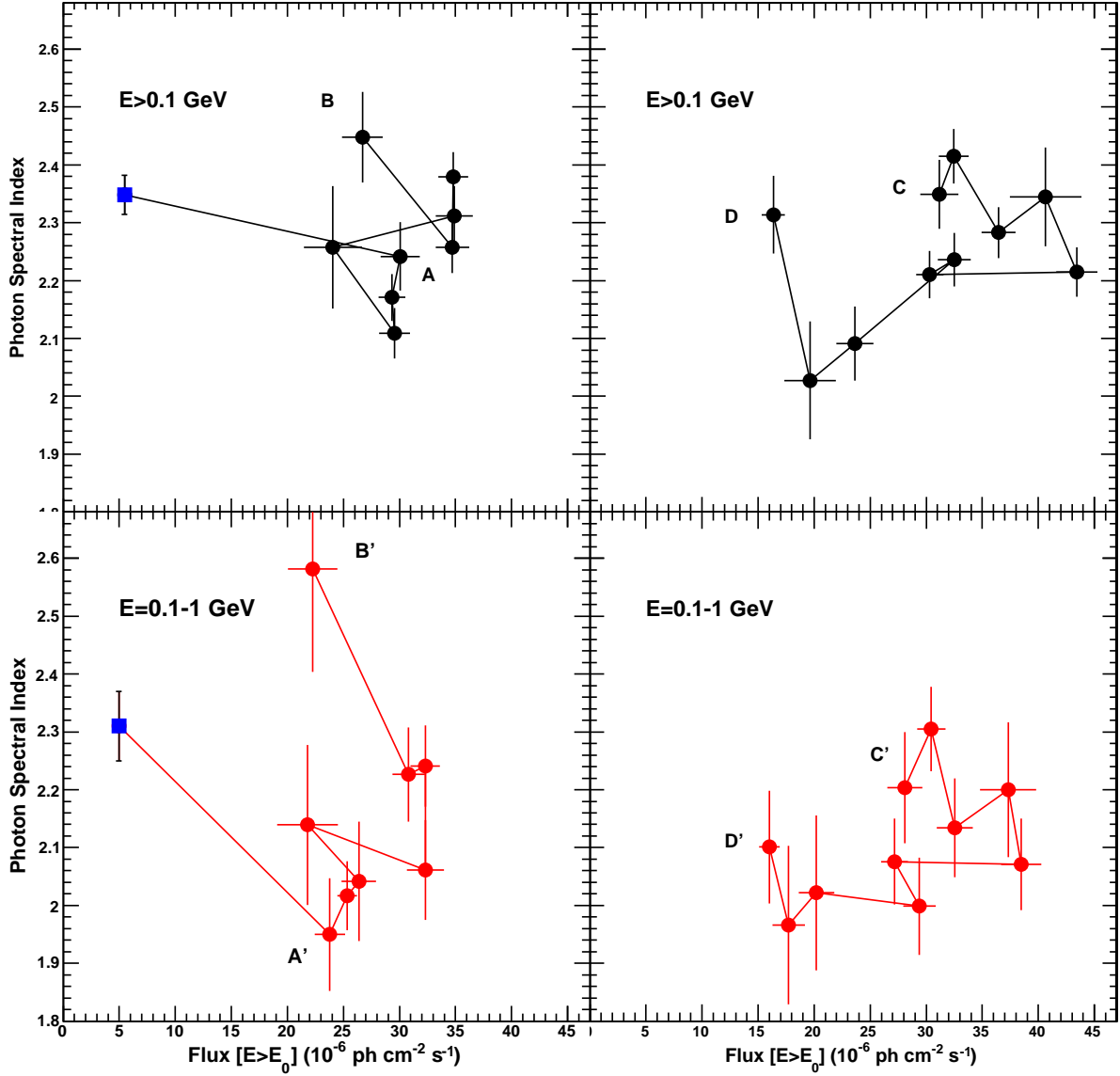}
\caption{$\Gamma$ vs. flux above the decorrelation energy for two different periods of time 
during the flare, obtained with a 6-hr binning. Photons with energy above 0.1$\:$GeV (top) or in the 0.1-1$\:$GeV range (bottom) were used. In the left panels, the blue points correspond to 4-day averages in the ``plateau'' period preceding the flare.}
 \label{fig:loops}
\end{figure}

The pronounced spectral evolution observed during the flare can be further investigated by plotting  $\Gamma$ versus the flux above $E_0=163\:$MeV. 
This is done, with a 6-hr binning in Figure \ref{fig:loops} 
for two consecutive time periods covering approximately the first and second half of the major flare and two different photon energy ranges: $E>0.1\:$GeV and $E=0.1-1\:$GeV. For orientation, the points associated with the earlier times have labels corresponding to those given in the third panel of Figure \ref{fig:light_curve_2}. In addition,  4-day averages obtained during the plateau period are displayed as blue squares.   
In contrast to the 2009 December-2010 April
flares for which no clear pattern was found in the data, a clockwise pattern is observed for the second period. The reduced
$\chi_r^2$  for a fit with a constant $\Gamma$ are 37.6/8 ($P=8.9\times10^{-6}$, $\sim 4.4\sigma$) and 35.4/8 ($P=2.3\times10^{-5}$, $\sim 4.2\sigma$) for the first and second periods, respectively (for the $E>0.1\:$GeV case). For the first period, a
flux increase by a factor of 4 is accompanied by an essentially constant $\Gamma$ (or one becoming weakly harder). 
This is followed by a clear hardening of the spectrum at constant flux, with
$\Gamma$ changing from $2.24\pm0.06$ to $2.11\pm0.04$ in 12$\:$hours.

 Such a  hard lag can also be observed in the top right-hand panel of Figure \ref{fig:loops}, where a hardening by 
$\Delta\:\Gamma=0.42\pm0.13$, associated with the {\sl decaying} stage of the flare, occurs over 2.25$\:$days. The
spectrum softens fairly quickly afterward, with $\Gamma$ changing from $2.0\pm0.1$ to $2.31\pm0.07$ in 6$\:$hours. 
This behavior may be driven by cooling. During the flare, the electron energy distributions may reflect the alternative dominance of acceleration and cooling processes. 
     
Restricting the analysis to the E=0.1-1$\:$GeV range (bottom panels in Figure \ref{fig:loops}) produces a pattern similar to the E$>$0.1$\:$GeV case for the second period, but somewhat different  for the first period, although the patterns are less clear due to larger statistical uncertainties. The rise in flux is accompanied by a pronounced hardening in this energy range followed by a state of essentially constant spectral hardness evolving into a slow softening. This behavior confirms
the conclusions obtained in the context of  Figure \ref{fig:light_curve_2}. Due to insufficient statistics, no clear pattern of the photon spectral index
above 1$\:$GeV vs flux can be observed with 6-hr time binning.

Figure \ref{fig:SED} shows the $\nu\:F_{\nu}$ distributions of the {\sl Fermi}-LAT data for the four different time periods 
delineated in Figure \ref{fig:light_curve_1}. These distributions have been fitted with BPL (solid), log-parabola function (dashed), and PLEC (dashed-dotted) functions.  
The parameters of the different fits can be found in the Table\ref{tab:funcs}.
 As the likelihood method does not provide an absolute goodness-of-fit measure, the $\chi^2$ of the $\nu\:F_{\nu}$ data points for the different functions have been calculated. For the pre-flare and plateau periods, both BPL and PLEC give fits of similar quality, while the log-parabola fit is notably worse. The PLEC function is preferred for the post-flare period.
 None of the tested functions provides 
a very good fit to the energy distribution in that period, which may be a result of the significant spectral evolution during the flare. The $\nu\:F_\nu$ spectra obtained over time intervals where the four subflares
alternatively dominate are consistent for the first three subflares in terms of
curvature, while a significantly harder spectrum is observed between MJD55520.0-55521.5.  In that interval, the PLEC fit gives  $E_\textit{cutoff}\;$=8.3$\pm$1.7$\:$GeV. A total of 10  photons with $E>10\:$GeV (out of 12 detected during the entire 5-day flare) were collected  in that 1.5-day time lapse (second panel of Figure \ref{fig:light_curve_2}).  The variation of $E_\textit{break}$ and $E_\textit{cutoff}$ with flux  are 
displayed in the inset of Figure \ref{fig:SED}. 
As already found during the 2009 December
and 2010 April flares, no strong evolution of either $E_\textit{break}$ or $E_\textit{cutoff}$ is found. 
$E_\textit{break}$  remains constant within a factor of $\approx2$ while the flux varies by  a factor of $\approx40$.

\section{Discussion}

\begin{figure}
  \centering
  \epsscale{1.0}
  \resizebox{14cm}{!}{\rotatebox[]{0}{\includegraphics{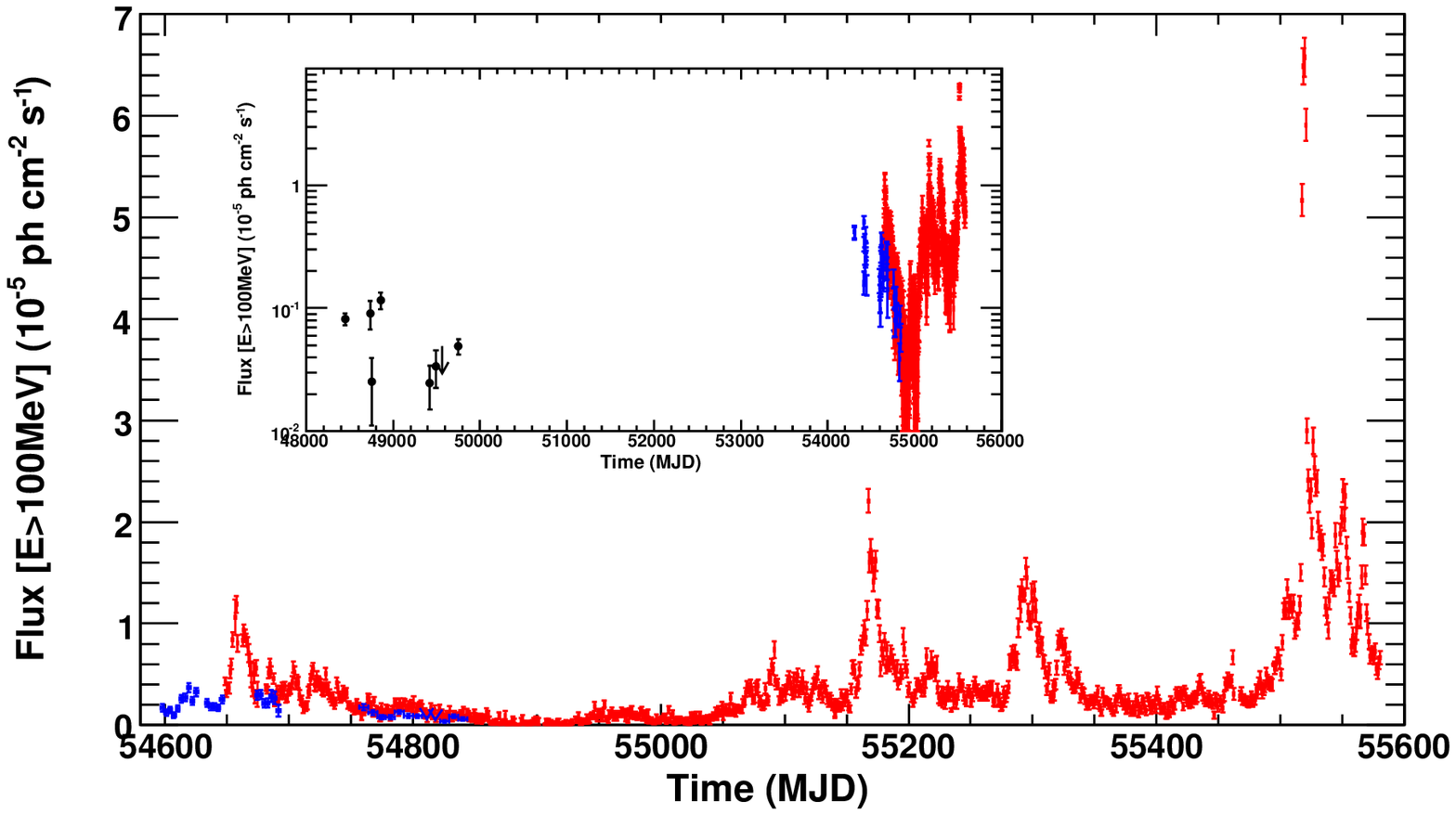}}}\\
  \resizebox{14cm}{!}{\rotatebox[]{0}{\includegraphics{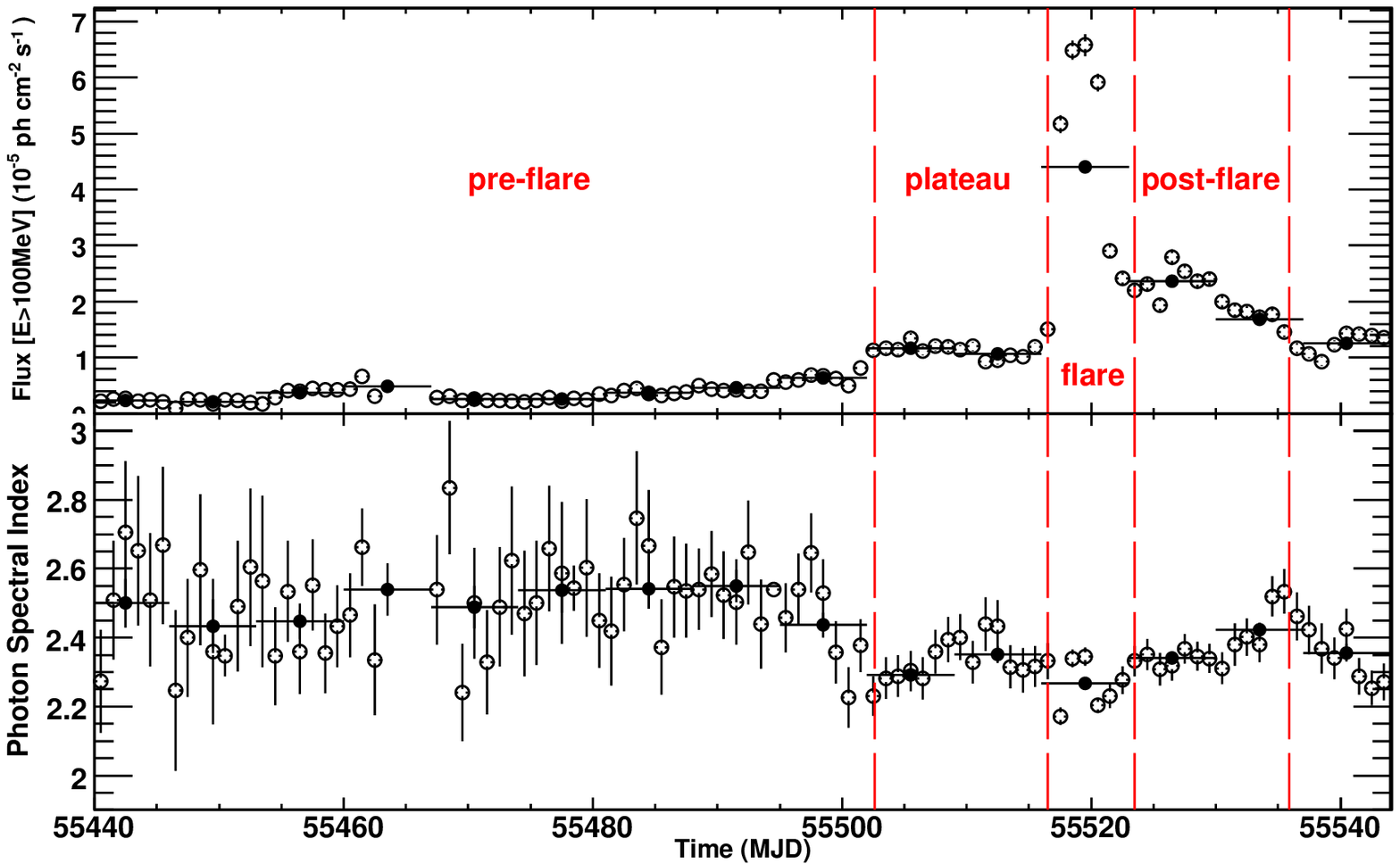}}}\\
  \caption{{\bf Top, main figure: Daily light curve of 3C454.3 measured with the Fermi-LAT since launch. Inset: Historical light curve. Black points are from EGRET \citep{Har99} and blue points are from AGILE \citep{Str10}.} Bottom: Light curve of the flux $F_{100}$ (top) and $\Gamma$ (bottom) for a 103-day period including both the slowly increasing
  flux phase, the plateau, the flare and the post-flare. The open and filled symbols correspond to daily and weekly
  averaged quantities respectively. Error bars are statistical only. }
  \label{fig:light_curve_1}
  \end{figure}

During its five-day outburst from 2010 November 17 to 21 (flare interval in Figure \ref{fig:light_curve_1}), 3C454.3 was the brightest GeV $\gamma$-ray source in the sky, with a flux $F_{100}=66\pm2$ on 2010 November 18--19. Prior to the flaring phase, the {\it Fermi}-LAT light curve displays a 13-day long flux plateau preceding the major outburst. The onset of the plateau is marked by a rapid ($<1$) day flux increase by a factor of $\approx 2$.  This feature appears to be a characteristic behavior indicating that 3C454.3 is about to flare, 
as noted in \cite{2010ApJ...721.1383A}. In the December 2009 flare, the 
plateau lasted for 6 days at a level of $F_{100}\approx10$ before flaring to 
a daily flux of  $\approx22$, 
while for the April 2010 outburst, it lasted for 7$\:$days at a level of $F_{100}\approx7$ before reaching a peak flux of  $\approx16$.
The spectrum hardens slightly from the pre-flare to the plateau preceding the giant flare. Spectral hardening and clustering of photons with $E>10\:$GeV is also seen in the decaying stage of the gamma-ray outburst at MJD55520.0-55521.5, which could point to the presence of radiating hadrons or $\gamma\gamma$-absorption effects. In the former case, protons require additional time to accelerate and cool while the $<1\:$GeV flux, if due to rapidly cooling electrons, would decline more rapidly. In the latter case, the emergence of the hard component could occur if the radiating plasma becomes optically thin to $\gamma\gamma$-absorption, either due to a larger bulk Lorentz factor or increased size of the radiating plasma.

\begin{figure}
\epsscale{0.8}
\plotone{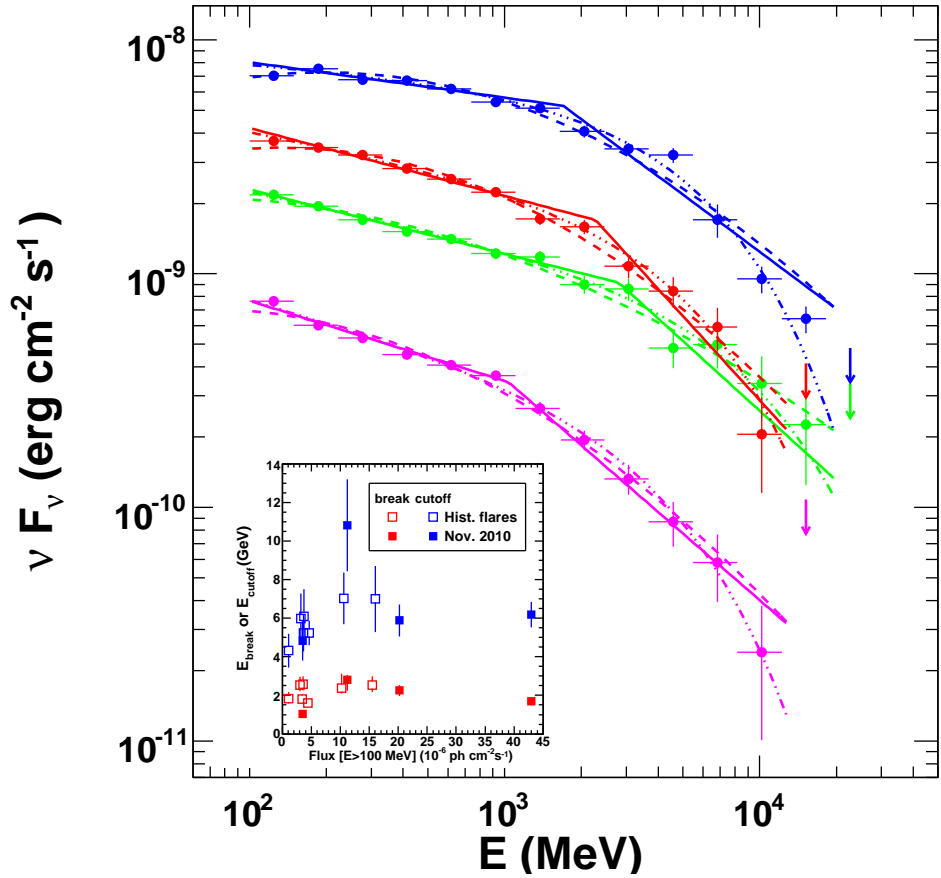}
\caption{Spectral energy ($\nu F_\nu$) distributions for four different time periods (pre-flare: magenta, plateau: green, flare: blue, post-flare: red), along with
the fitted BPL (solid), log-parabola (dashed) and PLEC (dashed-3dotted) functions. The inset displays $E_{\rm break}$ (red) and $E_{\rm cutoff}$ (blue) as a function of flux for the different periods considered here (filled symbols) and for historical flares (open symbols).}
\label{fig:SED}
\end{figure}

The features of the giant flare can be compared to those of the two earlier, fainter flares (December 2009 and April 2010) that have been carefully investigated in the LAT energy band (the different observation mode used during most of the July 2008 flare provided poorer-quality data). The overall light curves show similarities (presence of a preflare plateau, main flare lasting a few days, several week-long fading period). The rise time $T_r$ of the November 2010 flare is about half that of the December 2009 flare (4.5$\:$hr vs 8.9$\:$hr). Whereas the latter showed indication of ``flickering'' activity on timescales as short as 3$\:$hours above 100$\:$MeV, this effect is not clearly present here, as demonstrated by similar 3-hr and 6-hr light curves in Figure \ref{fig:light_curve_1}. For the first time, a significant temporary hardening of the spectrum leading to $\Gamma\simeq$2.1 has been observed for 3C454.3. Note that in the first LAT AGN catalog \citep{1LAC}, less than 2\% of FSRQs are found with 11-month averaged $\Gamma<\-$2.1. The moderately hard spectrum during the large luminosity flare deviates from the trend seen in the blazar divide \citep{Ghi09}, where the most $\gamma$-ray luminous blazars generally have $\Gamma\gtrsim2.5$. Despite the overall spectral variation, the energy cutoff remains essentially unchanged as observed in earlier flares (Figure \ref{fig:SED}). Interestingly, several-day long spectral variations are also observed  {\sl during the postflare period}  (beyond MJD55524 in Figure \ref{fig:light_curve_1}). No such effect was found in previous flares despite sufficient measurement statistical accuracy. Significant spectral hardening at the end of the main flare was not seen either.

The minimum Doppler factor $\delta_{min}$ can be numerically evaluated from $\gamma\gamma$-opacity constraints. From Swift-XRT public data, the total energy flux in the 2--10 keV range is $0.8\times10^{-10}\:\rm{erg}\:\rm{cm}^{-2}\:\rm{s}^{-1}$ and the photon-number index is 1.70. Correlated X-ray and GeV variability supports the assumption that the $\gamma$-rays are made co-spatially with the X-rays. At the time that the 31-GeV photon was observed, $t_{var}$=0.4 d (taken as $\ln(2)\times\:T_f$ in the fading phase of the flare), giving $\delta_{min}=16$, which is somewhat larger than the value of $\delta_{min}=13$ found by \citet{2010ApJ...721.1383A}. The estimated comoving size of the emission region is $R^\prime=c\:t_{var}\:\delta_{min}/(1+z)\:\approx\:3\times10^{15}$ cm. For the 31-GeV photon, the optical depth to pair production by the  extragalactic background light (EBL) is $\approx$1 for the high-EBL model of \citet{Ste06}, so no absorption constraints are provided by these data.

An upper limit on the optical depth $\tau_{\gamma\gamma}(E_{max})\approx2$ arising from the interaction of $\gamma$ rays with broad-line-region (BLR) photons can be obtained by comparing the flux measured at E$_{max}$=31$\:$GeV, with the flux extrapolated from lower energy. Assuming a BLR luminosity of 3$\times$10$^{45}$ erg s$^{-1}$\citep{Pia05} and adopting a BLR size $r_{BLR}\approx6\times10^{17}$ cm $\approx0.2$ pc from reverberation mapping \citep{Kas07,Bon10}, we calculate $\tau_{\gamma\gamma}(z_{em})$, where $z_{em}$ is the distance of the emitting blob from the black hole \citep[following][]{Rei07}. Assuming that the BLR clouds are distributed between 0.01$\:$pc and $r_{BLR}$, the condition $\tau_{\gamma\gamma}(E_{max})=2$ is satisfied for $z_{em}\simeq\:0.14\:$pc. Therefore the emission region must have been located either close to the outer boundary of the BLR or beyond at the time of emission of the 31-GeV photon.

The asymmetry of the time profiles derived for the subflares can be produced by acceleration/radiative losses or light-travel effects in quasi-spherical emission regions \citep{Sik01,Der04}. Assuming a spherical geometry of the emission region, then $r<2c\:\Gamma^2\:t_{var}/(1+z)\approx0.1\:$pc for a bulk Lorentz factor $\Gamma=\delta_{min}=16$ and a variability time scale $t_{var}\simeq0.4\:$d. If the jet opening angle $\theta_j\ll1/\delta_{min}\simeq3^\circ$, then this estimate can be compatible with the location estimated above.

The 3-hr peak $F_{100}$ is 85$\pm$5, corresponding to an apparent isotropic $\gamma$-ray luminosity $L_\gamma=10^{50}L_{50}$ erg s$^{-1}$ with $L_{50}=2.1\pm0.2$, assuming a spectral shape for the flare as given in Table$\:$1. This is $\approx4$ times the luminosity of PKS$\:$1622$-$297 ($L_{50}\simeq0.5$ with the current cosmological model) during its 1995 flare \citep{Mat97}, making this the largest $\gamma$-ray luminosity ever observed for a blazar. (\cite{Fos11} derived a luminosity of $L_{50}\approx3.0$ during this flare by assuming a power-law shape and considering a short 4.7$\:$ks time interval where $F_{100}\gtrsim100$.) During the 5-day flare, $F_{100}=43\pm1$ implies $L_{50}\simeq1.0$. The black hole mass for 3C454.3 is estimated to be in the range of (0.5--$4)\times10^9\:M_\odot$\citep{Bon10,Gu01}, so $L_{\rm Edd}\approx$(0.6--5)$\times10^{47}$ erg s$^{-1}$. In order for the time-averaged flare luminosity to be lower than $L_{\rm Edd}$, $\theta_j\lesssim2^\circ$--$6^\circ$, and a beaming factor $(1-cos\theta_j)^{-1}\gtrsim200$--1700 is implied. For highly efficient $\gamma$-ray production, the absolute jet power is comparable to the disk luminosity of $L_d$ =6.75$\times10^{46}\:\rm{erg}\:\rm{s}^{-1}$ estimated in \cite{Bon10}.

This flaring episode marks a record among AGNs and, indeed, all non-GRB sources for its rate of change in apparent luminosity, $L_\gamma/\Delta t$. Using a 6-hour variability time scale, then $L_\gamma/\Delta t\simeq10^{50}L_{50}$ erg s$^{-1}/(10^4\:t_4$ s) $\simeq10^{46}L_{50}/t_4\:\rm{erg}\:\rm{s}^{-2}$, noting that $6$ hr$/(1+z)\simeq10^4$ s. By comparison, the giant flares of PKS$\:$2155$-$304 \citep{Aha07} reached only $\lesssim10^{47}\:\rm{erg}$ s$^{-1}/300\:$s $\approx3\times10^{44}\:\rm{erg}$ s$^{-2}$ (in the TeV regime). This value now greatly exceeds $L_{\rm Edd}/(R_{\rm S}/c)\approx1.3\times10^{43}\:\rm{erg}$ s$^{-2}$ given by the ratio of the Eddington luminosity and the light crossing time across the Schwarzschild radius of a black hole, and strongly violates optically thin, Eddington-limited accretion scenarios \citep{Ell74}, showing that such a condition is unlikely to apply to the highly asymmetric disk/jet system of a blazar.

\acknowledgments 


{\bf We thank the AGILE team for providing their published data points.}

The \textit{Fermi}-LAT Collaboration acknowledges
support from a number of agencies and institutes
for both development and the operation of the
LAT as well as scientific data analysis. These include
NASA and DOE in the United States, CEA/Irfu and
IN2P3/CNRS in France, ASI and INFN in Italy, MEXT,
KEK, and JAXA in Japan, and the K. A. Wallenberg
Foundation, the Swedish Research Council and the National
Space Board in Sweden. Additional support from
INAF in Italy and CNES in France for science analysis during the operations
phase is also gratefully acknowledged.


\clearpage

\clearpage

\clearpage

\clearpage
{\footnotesize 
\begin{table}
\noindent \begin{tabular}{|c|c|c|c|c|c|c|c|}
\hline
period & $F_{100}$ & Luminosity & $\Gamma_1$ & $\Gamma_2$ &  E$_\textit{break}$ & $\Delta L $ & $\chi^{2}_r$ \\
 & (10$^{-6}\:$ph cm$^{-2}$s$^{-1})$ & (10$^{48}\:$erg s$^{-1}$) &  & & (GeV) & &\\ 
\hline
1 & 3.52$\pm$0.08 & 7.8  & 2.34$\pm$0.02 & 2.95$\pm$0.07 & 1.0$_{-0.1}^{+0.1}$ &  $-$31.6 & 6.3/8\\
\hline
2 & 11.2$\pm$0.2 & 26.3 & 2.28$\pm$0.02 & 3.00$\pm$0.10 & 2.8$_{-0.6}^{+0.3}$ &  $-$18.1 & 6.5/9 \\
\hline
3 &  43.0$\pm$0.6 & 105.8 & 2.15$\pm$0.01 & 2.81$\pm$0.05 & 1.7$_{-0.2}^{+0.1}$ & $-$74.5 & 45.9/9\\
\hline
4 &  20.2$\pm$0.3 & 45.5 & 2.29$\pm$0.02 & 3.20$\pm$0.10 & 2.3$_{-0.3}^{+0.3}$ & $-$44.4 & 16.6/8\\
\hline
\hline
period & $F_{100}$ & Luminosity & $\alpha$ & $\beta$ & - & $\Delta L $ & $\chi^{2}_r$   \\
 & (10$^{-6}\:$ph cm$^{-2}$s$^{-1})$ & (10$^{48}\:$erg s$^{-1}$) & & &  & & \\
\hline
1 & 3.45$\pm$0.07 & 7.7 & 2.61$\pm$0.03 &  0.11$\pm$0.01 & - & $-$26.3 & 19.6/9 \\
\hline
2 & 10.9$\pm$0.3  & 26.6 & 2.39$\pm$0.02 & 0.06$\pm$0.01 & -  & $-$13.8 & 12.5/10 \\
\hline
3 & 41.7$\pm$0.7 & 103.5 & 2.36$\pm$0.02 & 0.11$\pm$0.01 & - & $-$73.7 & 43.9/10  \\
\hline
4 & 19.1$\pm$0.4 & 44.4 & 2.49$\pm$0.02 & 0.12$\pm$0.01 & - & $-$37.4 & 13.0/9 \\
\hline
\hline
period &  $F_{100}$  & Luminosity & $\Gamma$ &  - & E$_\textit{cutoff}$ & $\Delta L $ & $\chi^{2}_r$ \\
& (10$^{-6}\:$ph cm$^{-2}$s$^{-1})$& (10$^{48}\:$erg s$^{-1}$) &  &  & (GeV) & &  \\
\hline
1 &3.5$\pm$ 0.1 & 7.7 & $2.30\pm0.04$ &  - & 5.0$\pm$1.0 & $-$24.1 & 9.4/9 \\
\hline
2 &11.1$\pm$0.4 & 26.0 & 2.23$\pm0.03$ &  - & 11.0$\pm$2.4 & $-$18.3 & 7.2/10 \\
\hline
3 &42.8$\pm$1.0 & 102.1 & 2.09$\pm0.02$ &  - & 6.2$\pm$0.7 & $-$85.5 & 22.8/10  \\
\hline
4 &20.0$\pm$0.5 & 44.8 & 2.21$\pm0.02$ &  - & 5.9$\pm$0.8 & $-$46.7 & 6.3/9 \\
\hline
\end{tabular}
\caption{Parameters of the BPL ($\Gamma_1$, $\Gamma_2$, E$_\textit{break}$), log-parabola ($\alpha$,  $\beta$)  and power-law+exponential cutoff 
($\Gamma$, E$_\textit{cutoff}$) functions fitted to the spectra for the different periods considered in Figure\ref{fig:SED}. 
$\Delta L$ represents the difference of the logarithm of the likelihood with respect to a single power-law fit,
 and  $\chi^{2}_r$ represents the reduced chi-squared of the $\nu\:F_\nu$ data points for the different functions.}
\label{tab:funcs}
\end{table}
}

\end{document}